\documentclass[runningheads]{llncs}
\usepackage{graphicx}
\usepackage{xcolor}
\usepackage{amsmath}
\usepackage{booktabs}
\usepackage{amsfonts}
\usepackage{chngcntr}
\usepackage{subfigure}
\usepackage{graphicx}
\usepackage[hyperfootnotes]{hyperref}

\begin{document}
\title{Probabilistic Radiomics: Ambiguous Diagnosis with Controllable Shape Analysis}
\titlerunning{Ambiguous Diagnosis with Probabilistic Radiomics}

\author{Jiancheng Yang\inst{1,2,3,}\thanks{These authors have contributed equally: J. Yang. and R. Fang.}  \and Rongyao Fang\inst{1,\star} \and Bingbing Ni\inst{1,2,3,}\thanks{Corresponding author: Bingbing Ni (nibingbing@sjtu.edu.cn).} \and \\Yamin Li\inst{1} \and Yi Xu\inst{1} \and Linguo Li\inst{1}}

\authorrunning{J. Yang et al.}
%
\institute{Shanghai Jiao Tong University, Shanghai, China\\
\and MoE Key Lab of Artificial Intelligence, AI Institute, Shanghai Jiao Tong University\\
\and Shanghai Institute for Advanced Communication and Data Science
\\\email{\{jekyll4168, lucas\_fang, nibingbing\}@sjtu.edu.cn}
}

\maketitle              
\begin{abstract}
Radiomics analysis has achieved great success in recent years. However, conventional Radiomics analysis suffers from insufficiently expressive hand-crafted features. Recently, emerging deep learning techniques, e.g., convolutional neural networks (CNNs), dominate recent research in Computer-Aided Diagnosis (CADx). Unfortunately, as black-box predictors, we argue that CNNs are ``diagnosing" voxels (or pixels), rather than lesions; in other words, visual saliency from a trained CNN is not necessarily concentrated on the lesions. On the other hand, classification in clinical applications suffers from inherent ambiguities: radiologists may produce diverse diagnosis on challenging cases. To this end, we propose a controllable and explainable {\em Probabilistic Radiomics} framework, by combining the Radiomics analysis and probabilistic deep learning. In our framework, 3D CNN feature is extracted upon lesion region only, then encoded into lesion representation, by a controllable Non-local Shape Analysis Module (NSAM) based on self-attention. Inspired from variational auto-encoders (VAEs),  an Ambiguity PriorNet is used to approximate the ambiguity distribution over human experts. The final diagnosis is obtained by combining the ambiguity prior sample and lesion representation, and the whole network named $DenseSharp^{+}$ is end-to-end trainable. We apply the proposed method on lung nodule diagnosis on LIDC-IDRI database to validate its effectiveness.

\end{abstract}

\setcounter{footnote}{0} 
\section{Introduction}

Medical images are more than pictures \cite{gillies2015Radiomics}. Mining hidden information using image analysis techniques is referred as {\em Radiomics} analysis, which raises numerous research attention in clinical decision making. 
Conventional Radiomics analysis follows the pipeline: 1) manual / automatic delineation of volumes of interest (VOIs); 2) image processing and feature extraction (e.g., SIFT, wavelet); 3) machine learning to associate features and target variables. These hand-craft features are named ``Radiomics". Though powerful and successful, emerging deep learning techniques indicate that hand-crafted features could be hardly comparable with end-to-end deep representations given enough data \cite{zhao2019toward}.

Deep learning\footnote{We refer to deep learning in a narrow sense, i.e., applying CNNs directly on the medical image analysis problems.} provides a strong alternative to learn representation from raw voxels (or pixels) in an end-to-end fashion. Convolutional neural networks (CNNs) have achieved great success in medical image analysis, though they are classifying {\bf voxels, rather than lesions}. In other words, there is no guarantee that black-box CNNs correctly learn evidence from lesions, especially with limited supervision. We illustrate several failures in Appendix Fig. \ref{fig:densenet-failure-cases}, by checking the Class Activation Maps (CAMs) \cite{zhou2016learning} from a 3D DenseNet \cite{huang2017densely,zhao2019toward} on lung nodule malignancy classification. These failures make the predictions given by CNNs unreliable. In contrast, Radiomics analysis is more controllable and transparent for users than black-box deep learning. 

On the other hand, classification in clinical applications suffers from inherent ambiguities; on challenging cases, experienced radiologists may produce diverse diagnosis. Though a ``ground truth'' to eliminate ambiguity could be obtained through a more sophisticated examination (e.g., biopsy) theoretically, this information may be unavailable from imaging only. Discriminative training procedure biases the model towards the mean values rather than ambiguity distribution.

To address these issues, we propose a controllable and explainable {\em Probabilistic Radiomics} framework. 
A $DenseSharp$ Network \cite{zhao20183d} is used as a backbone, which is a multi-task 3D CNN on learning classification and segmentation developed from 3D DenseNet \cite{huang2017densely,zhao2019toward}. Point clouds, named {\em feature clouds}, extracted from manual-labeled or predicted VOIs on CNN feature maps are regarded as lesion representations. To enable non-local shape analysis, we further introduce self-attention \cite{vaswani2017attention,yang2019modeling} to learn representations from the feature clouds. To capture label ambiguity, an Ambiguity PriorNet is used to approximate the ambiguity distribution over expert labels, inspired by Variational Auto-Encoders (VAEs) \cite{kingma2013auto}. By combining the ambiguity prior sample and lesion representation, the final decision is controllable (by lesion VOI) and probabilistic, which mimics the decision process of human radiologists. Please refer to Appendix Fig. \ref{fig:pipline_compare} for comparison among conventional Radiomics analysis, deep learning and Probabilistic Radiomics. On LIDC-IDRI \cite{armato2011lung} database, we validate the effectiveness of our methodology on lung nodule characterization from CT scans.

The key contributions of this paper are threefold: 1) We propose a novel viewpoint to regard deep representations from lesions on medical images as point clouds (i.e., feature clouds), and develop a Non-local Shape Analysis Module (NSAM) to end-to-end learn representations from  feature clouds (rather than voxels); 2) We explicitly model the diagnosis ambiguity within a probabilistic and controllable approach, which mimics the decision process of human radiologists; 3) The whole network named $DenseSharp^{+}$ is end-to-end trainable.

\section{Materials and Methods}

\subsection{Task and Dataset} \label{sec:dataset}
Lung cancer is the leading cause of cancer-related mortality worldwide. Early diagnosis of lung cancer with LDCT is an effective way to reduce the related death. In this study, we address the lung nodule malignancy classification problem to explore the performance of the proposed Probabilistic Radiomics method.

We use LIDC-IDRI \cite{armato2011lung} dataset, one of the largest publicly available databases for lung cancer screening. There are 2,635 nodules from 1,018 CT scans in the dataset, where nodules with diameters $\geq$ 3mm are annotated by at most 4 radiologists. For malignancy classification, rating mode ranges from ``1" (highly benign) to ``5" (highly malignant), while ``3" means undefined / uncertain rating. Besides, each radiologist delineates a VOI for a lesion. Empirically, the malignancy labels and segmentation VOIs are diverse for many instances in the dataset. Prior studies \cite{hussein2017risk,zhu2017deeplung} define a unique binary label for each instance by voting, we instead treat these labels with \textbf{ambiguity}, with all the \textbf{5 classes}. We called the whole dataset with 2,635 nodules as $HighAmbig$ (high ambiguous) dataset. To fairly compare the model performance, a $LowAmbig$ (low ambiguous) dataset is constructed, with a similar nodule inclusion criteria to prior studies \cite{hussein2017risk,zhu2017deeplung}: 1) the CT slice thickness $\leq$ 3mm, 2) annotated by at least 3 radiologists, and 3) the average rating $\neq$ ``3". The remaining nodules with average ratings $\leq$ ``3" are defined as benign, or malignant otherwise, resulting in 656 benign and 527 malignant.

We pre-process the data as follows: CT are resampled into $1mm \times 1mm \times 1mm$. The voxel intensity is normalized to $[-1,1)$ from the Hounsfield unit (HU), by $I= \lfloor \frac{I_{HU}+1024}{400+1024}\times 255\rfloor /128-1$. Each data sample is a voxel with a size of $32mm \times 32mm \times 32mm$. For simplicity, only single-scale inputs are used.

\subsection{Non-local Shape Analysis Module (NSAM)} \label{sec:nsam}

In our study, we use a CNN (DenseSharp \cite{zhao20183d} specifically) for extracting representations of nodules. Instead of a typical Global Pooling to derive the final classification, we use the lesion VOIs (manually annotated / automatically predicted) to crop the lesion features into point clouds \cite{yang2019modeling}, namely {\em feature clouds},  for subsequent processing. Inspired by self-attention transformer \cite{vaswani2017attention,yang2019modeling}, we develop a Non-local Shape Analysis Module (NSAM) to consume the feature clouds. 

Define $X \in \mathbb{R}^{N\times c}$ as a feature cloud, $X$ is a permutation-invariant and size-varying set. We figure out that self-attention is well suitable for set; besides, it enables non-local representation learning. We use scaled dot-product attention,

\begin{equation} \label{eq:attention}
    \mathit{Attn}(X) = \mathit{softmax}(XX^T/\sqrt{c})\cdot \sigma(X),
\end{equation}
where $\sigma$ is an activation function (e.g., ELU in our study).

Multi-head attention \cite{vaswani2017attention} is proved to be effective in attention  mechanism, where a scaled dot-product attention is applied multiple times on linear transformed input with various weights. The \textit{NSAM} is a variant of multi-head attention, by sharing the linear transformation weights in the $K, Q, V$-formation \cite{vaswani2017attention}. Define $g$ as the number of heads and $c_g=c/g$, the inputs are transformed by the weight $W_g \in \mathbb{R}^{c \times c_g}$ multiple times, before feeding into a scaled dot-product attention module. We further use skip connections \cite{he2016deep} to ease the optimization. 

\begin{equation}
    \mathit{NSAM}(X) = \mathit{concat}\{\mathit{Attn}(X_i)|X_i=XW_i\}_{i=1,..,g})+X.
\end{equation}

The whole shape analysis module is a stack of $L$-layer \textit{NSAM} ($L=3$, $c=256$ in this study). The features are subsequently fed into a global average pooling with multi-layer perceptron to obtain a single representation for a lesion VOI.

\subsection{Ambiguity PriorNet}
To deal with the ambiguous labels, we model the final decision as ambiguity prior distribution over the human experts. Inspired from Variational Auto-Encoders (VAEs) \cite{kingma2013auto}, a probabilistic module with a similar structure as 3D DenseNet backbone, named Ambiguity PriorNet (APN), is introduced to model the probabilistic component. APN produces $(\mu, \sigma)$, which controls a Gaussian distribution $N(\mu, \sigma)$ to serve as the ambiguity prior on malignancy labels and segmentation for human experts. To enable the gradient back-propagation, a reparameterization trick \cite{kingma2013auto} is applied to draw a prior sample $\mathit{f_{Ambig}}$ from $N(\mu, \sigma)$.

\begin{equation}
    \mathit{f_{Ambig}}(x) = \sigma x+\mu, x\in N(0,1).
\end{equation}

In subsequent modules, the prior sample $\mathit{f_{Ambig}}$ is concatenated with lesion representations to produce ambiguous malignancy labels and segmentation.

\subsection{$DenseSharp^{+}$ Network Architecture}

\begin{figure}[!htb]
 \centering
 \includegraphics[width=12cm]{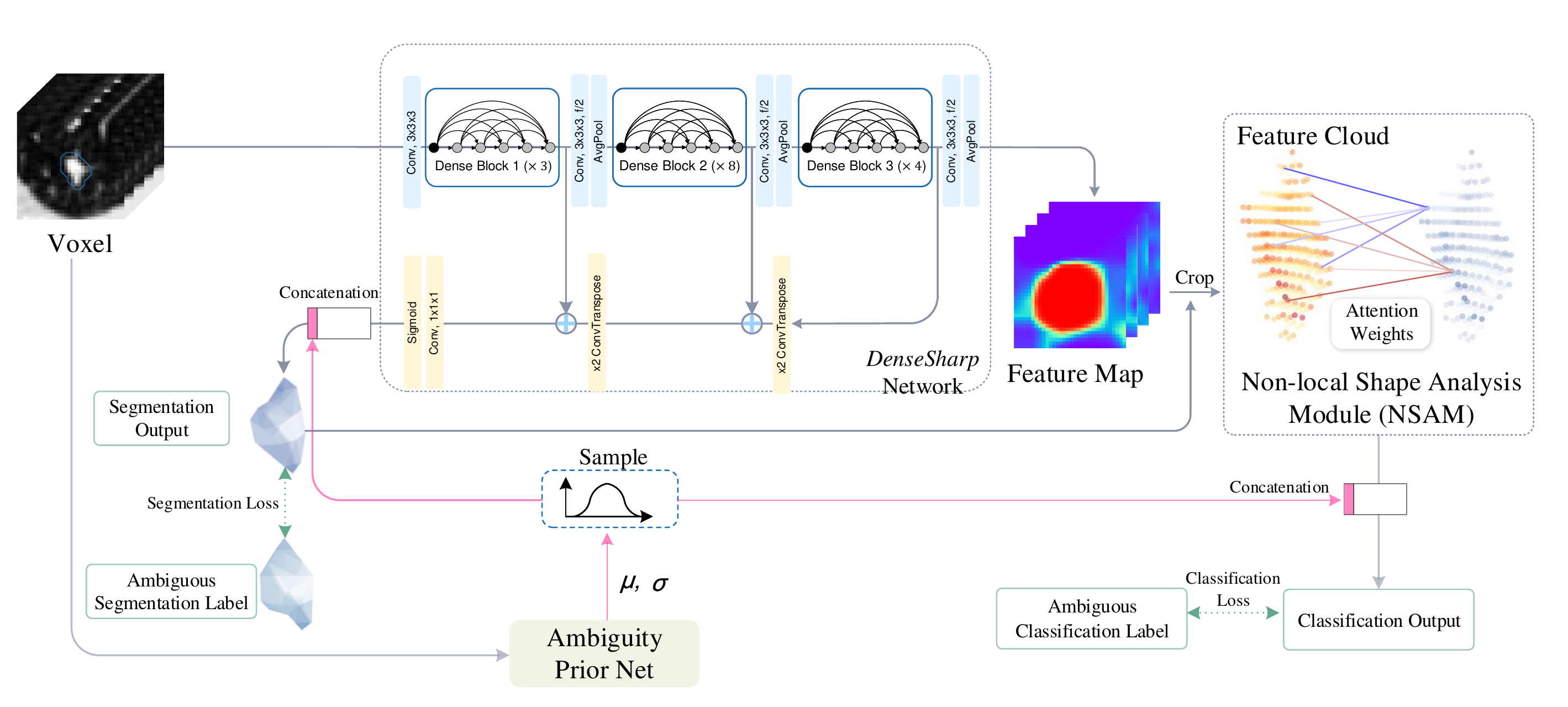}

 \caption{{\bf $DenseSharp^{+}$ Network Architecture}. A $DenseSharp^+$ Network is mainly a $DenseSharp$ Network followed by a Non-local Shape Analysis Module (NSAM). $DenseSharp$ is a deep 3D CNN based on DenseNet, with a classification head and segmentation head for multi-task learning. We use the feature maps from the classification head, cropped by manual / automatic segmentation, as {\em feature clouds}, rather than the raw feature maps, for the subsequent NSAM to consume. The NSAM use self-attention to associate non-local spatial information. An Ambiguity PriorNet conditional on the voxel inputs produces prior samples, which is concatenated with the classification and segmentation head to make their outputs probabilistic. Note the whole $DenseSharp^+$ Network is end-to-end trainable, with multi-task classification and segmentation loss.}
 \label{fig:model}
\end{figure}

The proposed $DenseSharp^{+}$ Network (Fig. \ref{fig:model}) is based on $DenseSharp$ Networks \cite{zhao20183d}, which is a multi-task 3D DenseNet \cite{huang2017densely,zhao2019toward} with classification and segmentation heads. The $DenseSharp$ Network uses a light-weight head for segmentation, which enables a top-down supervision for learning where the lesions are. At each resolution level ($32\times32\times32$, $16\times16\times16$ and $8\times8\times8$), dense blocks with 3D convolution and Batch Normalization \cite{ioffe2015batch} are repeated $[3,8,4]$ times before each down-sampling. Bottleneck ($B=4$), compression ($C=2$) and growth rate $k=32$ are used following the setting in the $DenseSharp$ paper \cite{zhao20183d}.

The feature maps outputted by the last convolution layer of classification head is upsampled (trilinear interpolation), and then cropped by the lesion segmentation into feature clouds which are consumed by NSAM (Sec \ref{sec:nsam}). Either manual or automatic segmentation by the segmentation head could be used as the lesion segmentation to generate the feature clouds. Although NSAM is able to process size-varying inputs, due to the GPU memory constraint, we sample up to $N^{\max}=1,024$ points from the feature cloud with sampling strategy $\Phi$. For the manual segmentation, the sampling strategy $\Phi$ is random sampling. For the predicted segmentation $\hat{y}_{seg}$, we first estimate the volume by $\hat{v}=\sum\hat{y}_{seg}$.
We then sample the $K=\lfloor\hat{v}\rfloor$ points with top-$K$ output scores from the segmentation head. If $N\leq N^{\max}$, all points in the feature cloud are selected. 

A DenseNet conditional on the voxel inputs (with a half parameter size of $DenseSharp$) is used as Ambiguity PriorNet (APN), which outputs 6-dimension prior samples to concatenate onto the classification and segmentation heads, to make their outputs probabilistic. Ideally, one prior sample encodes one ``human expert'', controlling the classification and segmentation results simultaneously.

\subsection{Training and Inference} \label{sec:training}


The $DenseSharp^+$ Networks is trained with two different schemes individually in order to better evaluate the probabilistic capability of the model. The first scheme trains on the $LowAmbig$ dataset (see Sec. \ref{sec:dataset}). This scheme denotes as $LowAmbig$ (low ambiguous) training scheme. The second scheme trains the model on the whole labeled dataset, which denotes as $HighAmbig$ (high ambiguous) training scheme. In both training schemes, unlike prior studies \cite{hussein2017risk,zhu2017deeplung} with a unique label on each voxel, we randomly select one of the four experts and the corresponding 5-class malignancy label and segmentation during training.

For training the multi-task neural networks, a cross entropy loss for classification and a dice loss for segmentation are used.
The loss weights for classification and segmentation are set as $1$ and $0.2$, respectively. Online data augmentation is applied on the voxels, including rotation, flipping and shifting within $[-1,1]$ on a random axis. We use Adam optimizer to train the whole $DenseSharp^+$ end-to-end with a batch size of 128 and a learning rate of $0.001$ for $150$ epochs.


For simplicity, feature maps from the $DenseSharp$ are cropped by predicted segmentation to feed into NSAM for training and inference. However, 
if the prediction segmentation volume is less than $10$, the model refuses to use it to classify the nodule. In this case, it is not counted in classification loss during training, and is ignored during the evaluation on classification.



\section{Experiments}

Our $DenseSharp^+$ Network is trained to classify ambiguous labels of $5$ malignancy modes from $4$ radiologists. $N$ prior samples ($N=10$ in our experiments) are obtained from the reparameterized conditional Gaussian distribution of the Ambiguous PriorNet. Hence, each tested voxel corresponds to $N$ 5-way outputs. In order to compare with prior studies quantitatively, the corresponding binary classification outputs are computed using Eq. \ref{eq:output}.

\begin{align}\label{eq:output}
\begin{split}
    &(p_1, p_2, p_4, p_5) = \frac{1}{N} \sum_{i=1}^{N}\mathrm{Softmax}(l_1^i, l_2^i, l_4^i, l_5^i),\\
    &p_b = p_1 + p_2, \hspace{10pt}p_m = p_4 + p_5,
\end{split}
\end{align}
where $l_1^i, l_2^i, l_4^i, l_5^i$ denote the $i^{th}$ logit outputs in the $N$ samples of mode $1$, $2$, $4$, and $5$ from 5-mode classification. Note mode $3$ is ignored in the evaluation since it defines ``uncertain'' diagnosis.

We evaluate the performance of all models via test AUC and accuracy on $LowAmbig$ LIDC-IDRI dataset (see Sec. \ref{sec:dataset}) with 5-fold cross validation method. It is worth noting that only $LowAmbig$ voxels are evaluated in all our experiments, since the binary labels for data in $HighAmbig$ are not trivially defined.

\begin{table}[!htb]
\centering
\caption{AUC and accuracy of DenseNet, $DenseSharp$, $DenseSharp^+$, and prior studies. The performance of our models is evaluated on $LowAmbig$ LIDC-IDRI \cite{armato2011lung} dataset (see Sec. \ref{sec:dataset}) with 5-fold cross validation.}
\begin{tabular}{lcc}
\toprule[1pt]
\multicolumn{1}{l}{Method}            & \multicolumn{1}{c}{AUC}    & \multicolumn{1}{c}{Accuracy (\%)}  \\ \cline{1-3}
\multicolumn{1}{l}{3D DPN \cite{zhu2017deeplung}}    & \multicolumn{1}{c}{-} & \multicolumn{1}{c}{88.28}         \\ 
\multicolumn{1}{l}{3D DPN ensemble \cite{zhu2017deeplung}}    & \multicolumn{1}{c}{-} & \multicolumn{1}{c}{90.44}         \\ 
 \cline{1-3}
 \multicolumn{1}{l}{3D CNN w. MTL \cite{hussein2017risk}}    & \multicolumn{1}{c}{-} & \multicolumn{1}{c}{80.08}         \\ 
\multicolumn{1}{l}{3D CNN w. sparse MTL \cite{hussein2017risk}}    & \multicolumn{1}{c}{-} & \multicolumn{1}{c}{91.26}         \\ 
 \cline{1-3}
\multicolumn{1}{l}{3D DenseNet (our implementation)}    & \multicolumn{1}{c}{0.9218} & \multicolumn{1}{c}{87.82}         \\ 
\multicolumn{1}{l}{$DenseSharp$ \cite{zhao20183d}} (our implementation) & \multicolumn{1}{c}{0.9393} & \multicolumn{1}{c}{89.26}        \\ 
\multicolumn{1}{l}{$DenseSharp^{+}$ (LowAmbig)} & \multicolumn{1}{c}{0.9480} & \multicolumn{1}{c}{90.87}         \\
\multicolumn{1}{l}{$DenseSharp^{+}$ (HighAmbig)} & \multicolumn{1}{c}{\textbf{0.9566}} & \multicolumn{1}{c}{\textbf{91.52}}         \\
\bottomrule[1pt]
\end{tabular}
\label{table:model-performance}
\end{table}

Table \ref{table:model-performance} shows the performance of our models and baselines\footnote{Note that all counterparts use (sightly) different evaluation protocols.}. It is noticeable that 3D DenseNet reveals a comparable performance with 3D DPN \cite{zhu2017deeplung}. The $DenseSharp^+$ network with $HighAmbig$ training scheme outperforms the one with $LowAmbig$ training scheme. The $HighAmbig$ $DenseSharp^+$ is trained on an ambiguous dataset with a larger scale, resulting in a better performance than that of $LowAmbig$ $DenseSharp^+$, which shows an excellent ability to learn from the ambiguous data distribution. The performance of $HighAmbig$ trained $DenseSharp^+$ is also better than 3D DPN ensemble \cite{zhu2017deeplung} and 3D CNN w. sparse MTL \cite{hussein2017risk}. Notably, compared with other methods, we adopt a coarser dataset pre-processing strategy and a simpler evaluation setting. For instance, both counterparts \cite{zhu2017deeplung,hussein2017risk} use 10-fold cross validation, with more training samples than 5-fold in our study. The 3D DPN \cite{zhu2017deeplung} only evaluates its performance on the overlapping nodules with LUNA16 dataset, which are easier to classify. The sparse MTL \cite{hussein2017risk} resamples voxels at a higher resolution (spacing of $0.5mm$), besides the CNN is pre-trained on large-scale video dataset, rather than randomly initialized. 

As for the segmentation output of $DenseSharp^+$, the average segmentation dice coefficient is $0.7625$ on $LowAmbig$ LIDC-IDRI with 5-fold cross validation. The segmentation output is of good quality with such a light-weight segmentation head. Due to the probabilistic segmentation output, $DenseSharp^+$ with automatic segmentation refuses to classify the nodules whose predicted volume is less than 10; $73$ nodules are refused by $HighAmbig$-trained $DenseSharp^+$.



\begin{figure}[!htb]
\vspace{-20px}
\centering
\includegraphics[width=10cm]{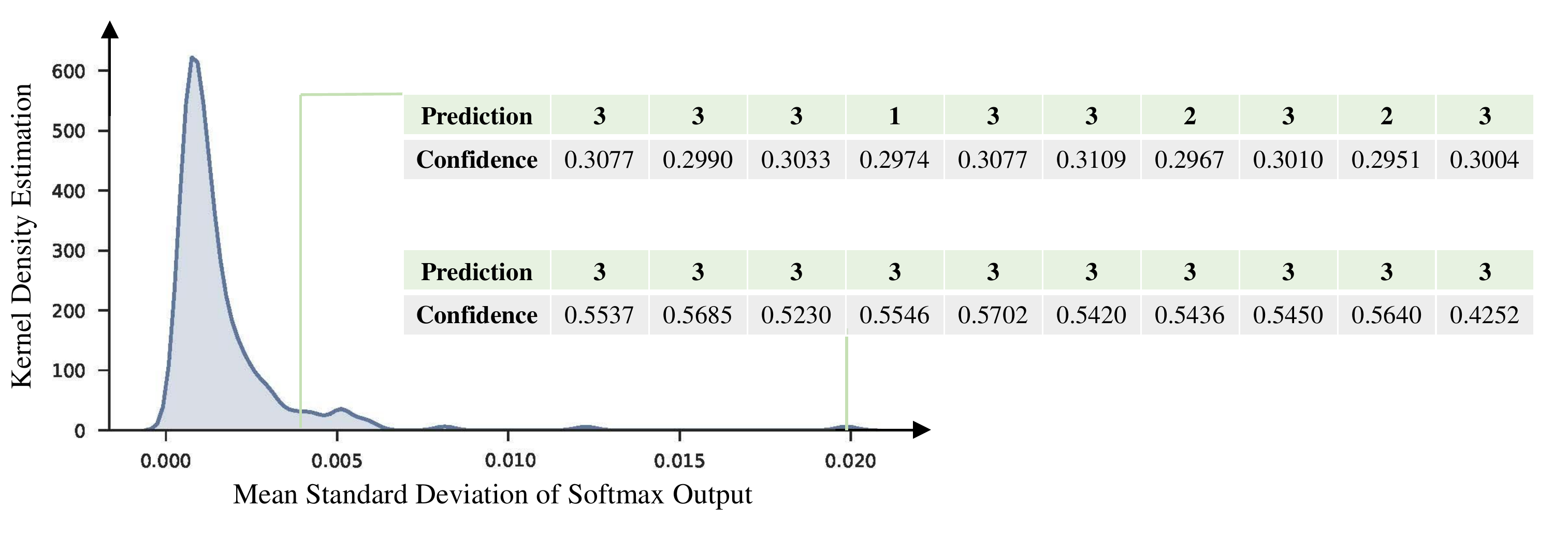}
\label{fig:quantification}
 
\caption{The diversity metric ($DIV$) distribution of all tested voxels. The two highlight examples show that the output of $DenseSharp^+$ model varies as the prior sample varies, thanks to its probabilistic property.}
\vspace{-20px}
\end{figure}

For further evaluation of probabilistic property of $DenseSharp^+$ model, we compute the mean standard deviation of softmax outputs as a diversity metric, derived from the softmax outputs of all the tested voxels (Eq. \ref{eq:msdso}),

\begin{equation}\label{eq:msdso}
    \mathrm{DIV} = \frac{1}{5}\sum_{i=1}^{5}\mathrm{Std}_{j=1...N}(p_{ij}),
\end{equation}
in which $p_{ij}$ is the softmax output of malignancy mode $i$ and $j^{th}$ sample of Guassian distribution from one voxel. $\mathrm{Std}(\cdot)$ is the standard deviation operation. The distribution of $DIV$ from all the tested voxels reflects the probabilistic output variance of $DenseSharp^+$ Networks. Figure \ref{fig:quantification} shows the $DIV$ distribution of all the tested voxels. The two highlight samples show that the classification predictions from the model mimic the ambiguous labels from different experts.

Moreover, thanks to the explicit modeling, only voxels in lesions are counted, the visual saliency maps produced by the $DenseSharp^+$ is highly calibrated with the nodules. Please refer to Appendix Fig. \ref{fig:cam-in-result} for illustration.

\section{Conclusion and Further Work}

In this study, a Probabilistic Radiomics framework is proposed, which is well-performing, controllable and explainable in Computer-Aided Diagnosis (CADx). The proposed method is more expressive than conventional Radiomics analysis, more controllable and explainable than conventional deep learning approaches. Moreover, we explicitly model the ambiguity of the classification with a probabilistic approach. However, there are still limitations to make the Probabilistic Radiomics an \textit{omics}-level approach  (e.g., genomics, proteomics, immunomics).

Compared to other ``omics" approaches, Radiomics is generally less reproducible \cite{gillies2015Radiomics}. Perturbations (e.g., rotations, different imaging parameters, adversarial attacks) on the images / point clouds \cite{yang2019adversarial} could introduce large variances to the outputs. Besides, the data-hungriness issue makes current MIC research a Sisyphean challenge; model learning on a certain task is non-trivial to transfer to another task. A more generalizable representation learning is the key to this problem, (probably) following a route of self-supervised learning and meta-learning. We will explore the robustness, transferability, and reproducibility of Probabilistic Radiomics in the future study.

\subsubsection{Acknowledgment.}
This work was supported by National Science Foundation of China (U1611461, 61502301, 61521062). This work was supported by SJTU-UCLA Joint Center for Machine Perception and Inference, China's Thousand Youth Talents Plan, STCSM 17511105401, 18DZ2270700 and MoE Key Lab of Artificial Intelligence, AI Institute, Shanghai Jiao Tong University, China. This work was also jointly supported by SJTU-Minivision joint research grant.
\bibliographystyle{splncs04}
\bibliography{ref}
%





\clearpage
\appendix
\counterwithin{figure}{section}
\section{Appendix Figures}
\vspace{-20px}

\begin{figure}[!htb]
  \centering
  \includegraphics[width=5cm]{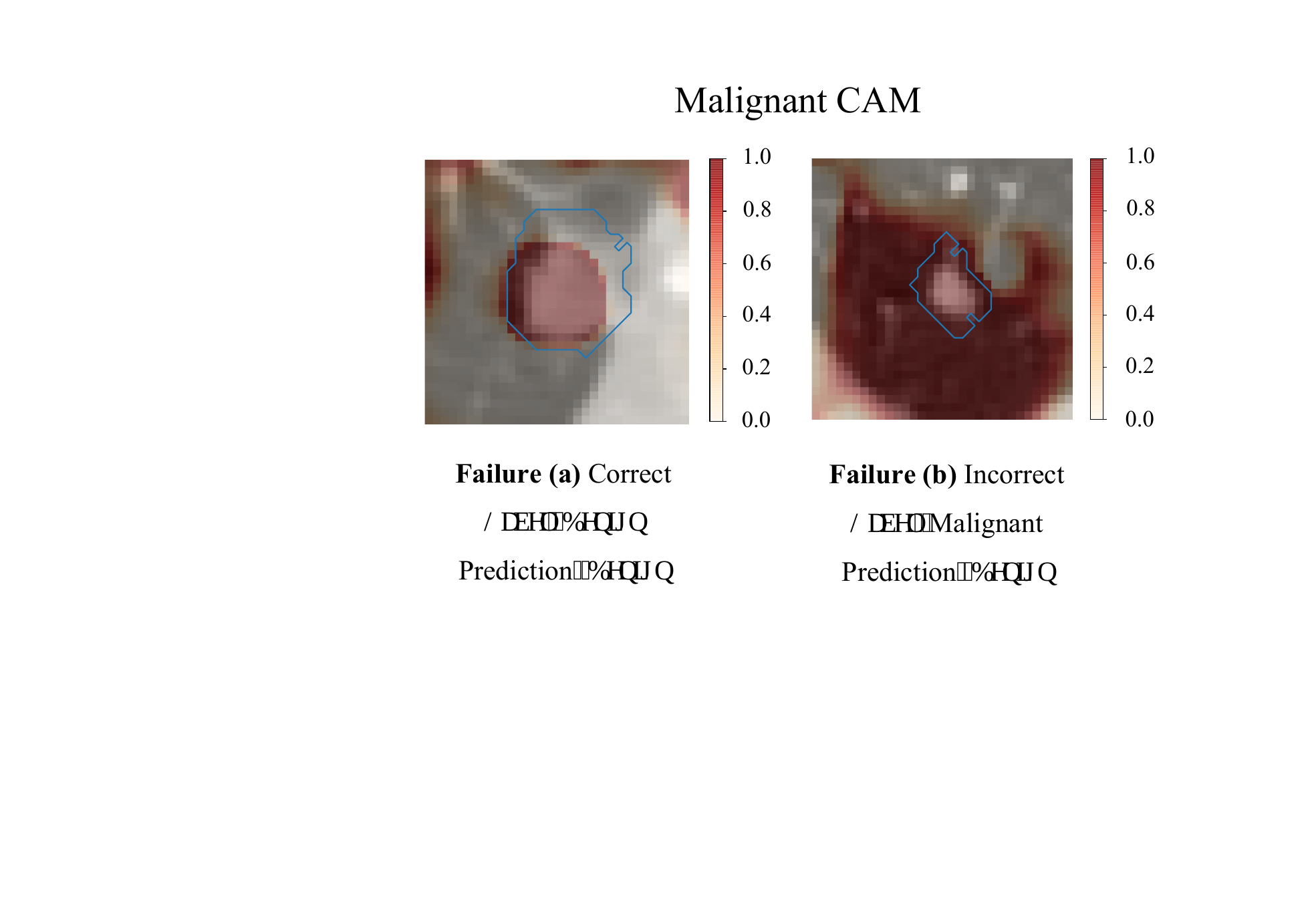}
  \caption{Two types of failures from a well-trained 3D DenseNet, visualized by CAM techniques. Only malignant CAMs on the central slices are depicted. The blue contours on each plot are manual segmentation of lesions by radiologists. The voxels with higher intensity are more malignant, and those with intensity $\leq 0.5$ are benign. For failure (a), the model predicts ``benign" on a benign nodule correctly. However, this ``correct" prediction comes from the prediction apart from lesions on voxels, which means the model uses incorrect evidences. For failure (b), the model outputs ``benign" on a malignant nodule incorrectly. Whereas, within the lesion voxels it is indeed predicted as malignant, indicating that the model performance could be boosted further if it uses correct evidences.}
  \label{fig:densenet-failure-cases}

  \vspace*{\floatsep}

  \includegraphics[width=10cm]{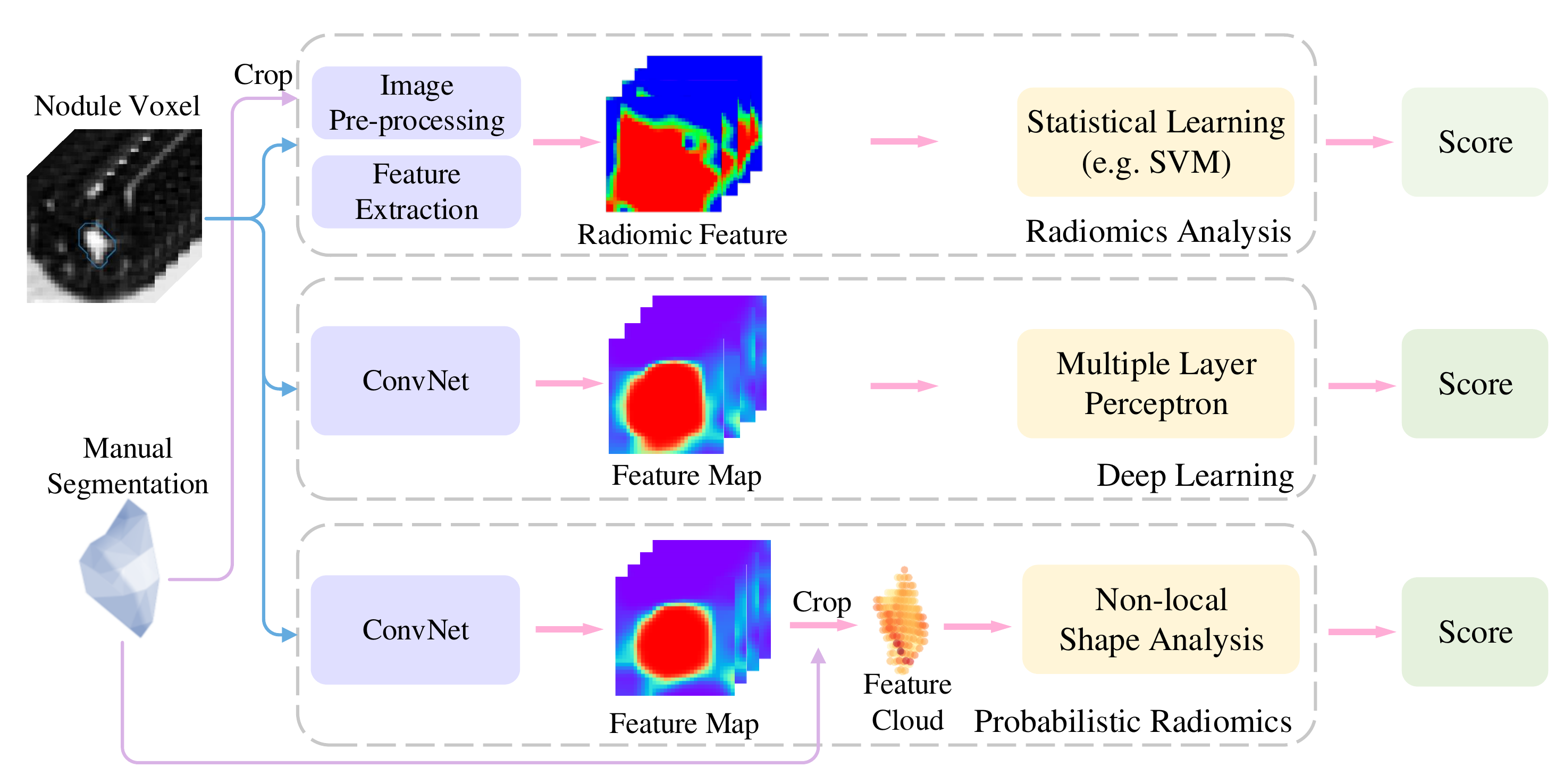}
  \caption{Comparison of conventional Radiomics analysis, deep learning, and our proposed Probabilistic Radiomics framework. Radiomics analysis (top) only responds to the user-delineated VOIs, while the hand-crafted features are pre-defined and not learnable. Conventional Deep learning (middle) learns expressive representations end-to-end from voxels of CT scans, however, it could possibly learn ``evidences" outside lesions, making its prediction unreliable and unexplainable. The proposed Probabilistic Radiomics framework (bottom) uses {\em feature clouds} (instead of voxels) for a final decision, which are CNN feature maps cropped by the automatic segmentation of lesions. The feature clouds are then consumed by a Non-local Shape Analysis Module (NSAM) based on self-attention for deeper representation. The proposed framework takes advantage of the expressiveness of deep learning and the controllability of Radiomics analysis, thus defining a {\em Probabilistic Radiomics}.}
  \label{fig:pipline_compare}
\end{figure}

\begin{figure}[!htb]
\centering
\includegraphics[width=12cm]{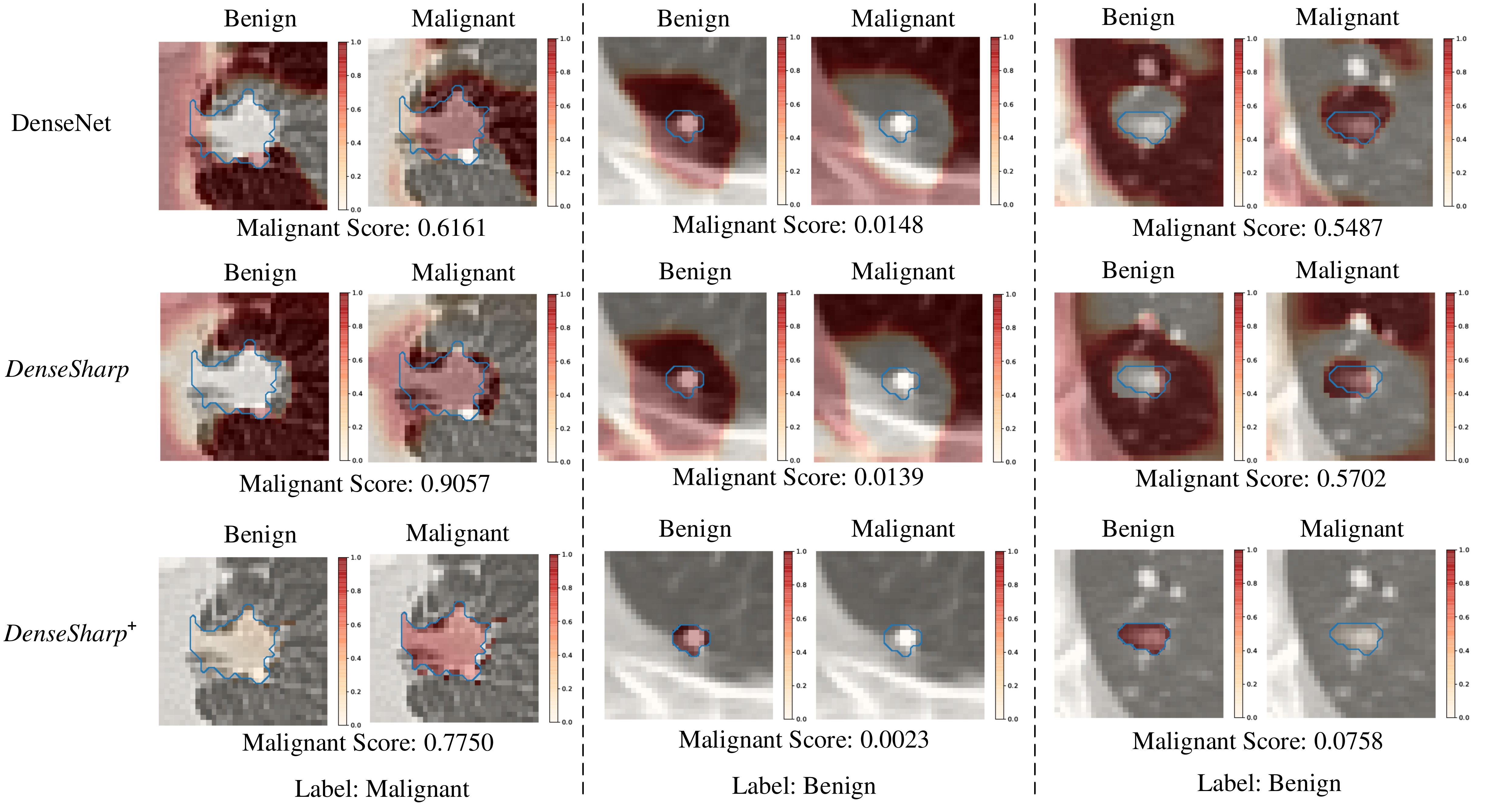}
\caption{Three nodule samples classified by DenseNet, $DenseSharp$, and $DenseSharp^{+}$, visualized by CAM techniques. As the benign and malignant CAMs have gone through a \textit{softmax}, the sum of benign CAM and malignant CAM in a corresponding voxel equals to $1$. The blue contours on each plot are manual segmentation of lesions. The "labels" are the classification by radiologists. The malignant scores are the possibilities of malignancy (predicted by models). The threshold of output score is $0.5$ (larger than $0.5$ classify as malignant and vice versa). As illustrated, the segmentation head of $DenseSharp$ helps the model better locate the lesions than DenseNet, making the CAM of $DenseSharp$ appears a more precise activation than that of DenseNet to the manual segmentation. In most cases, DenseNet and $DenseSharp$ models not only activate the features in lesions' locations, but also activate the locations in the background, which not precisely utilizes the features of lesions themselves (the "correct evidences"). In some other cases, the two models face the two failures described in Fig. \ref{fig:densenet-failure-cases}, making their classification incorrect or lack of interpretability. $DenseSharp^+$ model only adapts the features upon lesions to classify the nodule, with better controllability and interpretability. }
\label{fig:cam-in-result}
\end{figure}


\end{document}